# Dynamics of Structural Inequality: Multidisciplinary Analysis of Meritocracy, Media Influence, Functionalism

Yasuko Kawahata [†]

Faculty of Sociology, Department of Media Sociology, Rikkyo University, 3-34-1 Nishi-Ikebukuro,Toshima-ku, Tokyo, 171-8501, JAPAN.
`ykawahata@rikkyo.ac.jp,kawahata.lab3@damp.tottori-u.ac.jp`

**Abstract:** This study analyzes the role of meritocracy, media influence, and scheduled theory from multiple perspectives as mechanisms that maintain inequality in social classes. Social inequality exists in complex forms in the educational, media, and political spheres. The study focuses on how inequality in society is structured and reproduced and how the theory of scheduled harmony justifies this.

## 1. Introduction

This study analyzes the role of meritocracy, media influence, and parameter scheduling theory from multiple perspectives as mechanisms that maintain inequality in social strata. In particular, it will examine in detail the reality of social inequality in terms of regional disparities in education and health, considering how the media shape and reinforce perceptions and stereotypes of social class. Through modeling with opinion dynamics, we will reveal how deeply rooted social inequalities are and the mechanisms of their reproduction, and then how scheduling rationalizes and maintains these disparities. This analysis will provide insight into how interactions between media and social classes contribute to the perpetuation of social inequalities, and hint at inferences about how scheduling generates structures that favor certain classes and groups. Finally, opinion dynamics and social systems simulations will be used to examine the mechanisms of inequality maintenance and their impact on social policy. We will also explore the process by which meritocracy creates conditions that favor certain social classes while marginalizing others, and analyze the impact of socioeconomic barriers on media literacy and information access. This multifaceted study will consider theoretical foundations and practical measures to promote change toward an equitable social structure. Prospects should also examine the perspective of the contribution of each social element to the whole as proposed by the theory of scheduled harmony, and explore how this in practice immobilizes social inequalities and marginalizes certain groups.

By applying opinion dynamics and integrating social, educational, and economic processes, this analysis provides a new perspective on the mechanisms that maintain inequality and their impact on overall social cohesion. We hope that this study will also be an element in furthering our understanding of the ways in which meritocracy and the theory of scheduled settlements shape and maintain social hierarchies. The analysis incorporates a novel approach using simulations of opinion dynamics for six social classes.

## 2. Previous Research

Although we will apply existing research on opinion dynamics with respect to this paper, there have been very old and diverse approaches in the social sciences. We would like to summarize them first as examples, and then move on to modeling, simulation, and discussion based on the hypotheses.

### 2.1 Case Studies of Discrimination, Prejudice, and Disparity Issues Related to Social Class

Piketty, T. (2014), Capital in the Twenty-First Century, analyzes long-term trends in economic inequality, noting that the concentration of income and wealth creates disparities between social classes. He examines in detail the imbalances between economic growth and capital returns, focusing on the mechanisms of wealth accumulation and distribution. Wright, E. O. (1997), in his book Classes, discusses how social class affects individual life opportunities and outcomes, examining changes in class structure and their impact on individual life paths; Gilens, M. (1999), Why Americans Hate Welfare" analyzes Americans' views of public support and how they are rooted in preconceived notions of class and race, and discusses the role of the media in reinforcing these notions; Kerbo, H. R. (2012)'s "Social Stratification and

Sayer, A. (2005), The Moral Significance of Class, provides a systematic treatment of theories of stratification, how stratification is measured, and various forms of inequality, providing material for understanding class and inequality from a global perspective. Class, examines in detail the impact of class on individual identity, self-esteem, and daily life, and argues that class is deeply related to individual and group moral sentiments beyond economic status; Adler, N. E., and Stewart, J. (2010), in Health Disparities across the Lifespan: Meaning, Methods, and Mechanisms, studies how health inequalities are affected by socioeconomic status, highlights how class differences affect health access and outcomes, and explores how the results can inform policymaking. Corak, M. (2013) in her article "Income Inequality, Equality of Opportunity, and Intergenerational Mobility" analyzes how income inequality affects the future socioeconomic status of children and The paper analyzes how income inequality affects children's future socioeconomic status and discusses in detail the impact of parental income on children's success. The issue of social class inequality has been the subject of extensive research, examining class-based disparities in a variety of areas, including education, health, employment, and housing. Putnam, R. D. (2015), Our Kids: The American Dream in Crisis, empirically explores how American stratification affects inequality in educational opportunity. It reveals how socioeconomic status affects children's academic achievement and social advancement; Wilkinson, R. G., and Pickett, K. (2009), The Spirit Level: Why More Equal Societies Almost Always Do Better" proposes the theory that many different social problems (health, crime, education, etc.) depend on the degree of social inequality. They argue that more equal societies are healthier societies, and Dorling, D. (2010), Injustice: Why Social Inequality Persists, provides a detailed analysis of five inequality-maintaining mechanisms (elitism, exclusionism, exclusion, scheduling, and shaming), which they call social inequality. Chetty, R., Hendren, N., Kline, P., and Saez, E. (2014) in their research paper "Where is the Land of Opportunity? Intergenerational Mobility in the United States," provides a detailed analysis of how geographic location in the United States affects class mobility. They point out that children with higher-income parents are more likely to have better educational and economic opportunities than children from lower-income families. Marmot, M. (2004), Status Syndrome: How Social Standing Affects Our Health and Longevity, a study of how social class affects health, found that income, education, and occupational status are closely related to health It shows that income, education, and occupational status are closely related to health. Sassen, S. (2014), Expulsions: Brutality and Complexity in the Global Economy, analyzes how the complex processes developing in the global economy increase disparities and exclude certain social classes. Acemoglu, D., and Robinson, J. A. (2012), Why Nations Fail: The Origins of Power, Prosperity, and Poverty, discusses how political and economic institutions shape the economic success and failure of nations, resulting in class disparities.

## 2.2 Social Class and the Media

Social class and the media have a complex relationship that influences each other. The media often serve to shape the image of social class and reinforce society's perceptions and stereotypes about class. In addition, access to and content of the media itself may vary by class. This article illuminates the different ways in which media shape and reinforce class perceptions. Perspectives on the intersection of media and class are important for future theory building, as we study this topic from multiple perspectives, including cultural capital, political identity, national affiliation, and stereotypes. Bourdieu, P. (1984), Distinction: A Social Critique of the Judgement of Taste, explores how tastes and consumer behavior reproduce class discrimination. It is argued that the media play an important role in this process; Couldry, N. (2000), The Place of Media Power: Pilgrims and Witnesses of the Media Age, analyzes how the media contribute to the construction of social class. The book analyzes how the media contributes to the construction of social class. It focuses specifically on the power dynamics between media events and social classes.Lamont, M., and Lareau, A. (1988), Cultural Capital: Allusions, Gaps and Glissandos in Recent Theoretical Developments, uses the concept of cultural capital to explore differences in media use and interpretation by class. Curran, J. (2011) in Media and Society discusses how media reflect and reinforce social identity and class structure. Davis, A. (2006), Public Relations Democracy: Public Relations, Politics and the Mass Media in Britain, discusses how the media interact with political processes and shape perceptions of class structure in Lamont, M., and Lareau, A. (1988), Cultural Capital: Allusions, Gaps and Glissandos in Recent Theoretical Developments, uses the concept of cultural capital to examine how class Skeggs, B. (2004), Class, Self, Culture, explores how class identity is expressed and used by the media Mihelj, S. (2011), Media Nations: Communicating Belonging and Exclusion in the Modern World, delves into the relationship between nation, media, and class, examining how media shape national identity through class. Tyler, I. (2008), "Chav Mum Chav Scum': Class Disgust in Contemporary Britain', studies how the media popularize class-related pejoratives.

## 2.3 How Media Reflect Social Identity and Class Structure

Many case studies on the ways in which the media reflect social identity and class structure have been conducted in the areas of sociology, communication studies, and cultural stud-ies. These studies explore how the media function as a means of shaping class consciousness, social identity, and cultural hegemony. Approaches range from theoretical frameworks to empirical analyses. Each of the literature is also an important cognitive psychology item in how class is played out and perceived through different media forms, such as television programming, news media, advertising, and digital communication. Gramsci, A. (1971)'s Selections from the Prison Notebooks develops a theory of how cultural hegemony and media shape class relations; Hall, S. (1973)'s Encoding and Decoding in Thompson, J. B. (1995), The Media and Modernity: A Social Theory of the Media" analyzes how the media relate to contemporary social structures and individual identity; in Golding, P., and Murdock, G. (2000), Culture, Communications, and Political Economy, discusses how the ownership structure of the media industry mirrors class structures; Bettie, J. (2003), Women without Class: Girls, Race, and Identity, studies the intersection of gender, race, and class, and specifically examines the influence of media on young women's Castells, M. (2004), The Power of Identity, positions media as a source of identity in the information age and describes identity transformations related to class structures. Castells, M. (2004), The Power of Identity. In Skeggs, B., and Wood, H. (2012), Reacting to Reality Television: Performance, Audience and Value, how reality television represents class structure and how it affects viewers is Investigating the impact of digital media on political identity and class activism in Fenton, N. (2016), Digital, Political, Radical. It provides a theoretical framework and empirical data on how media shape, communicate, and reinforce notions of class. Another important item to study the role of the media from a social class perspective would be how class structures are maintained and changed in Couldry, N. (2000), The Place of Media Power: Pilgrims and Witnesses of the Media Age, Bourdieu, P. (1984), Distinction: A Social Critique of the Judgement of Taste, explores how taste shapes social class in the context of class discrimination. It explores how taste shapes social class and analyzes how the media reflect this; Curran, J. (2002), Media and Power, discusses how the media reflect power structures and what role they play in social class; and Bourdieu, P. (1984), Taste: A Social Critique of the Judgement of Taste, explores the role of taste in class discrimination and how it is reflected in the media. Davis, A. (2019), Underdog Politics: The Minority Party in the U.S. House of Representatives, studies how political media influence debates about class. Herman, E. S ., and Chomsky, N. (1988), Manufacturing Consent: The Political Economy of the Mass Media, addresses how the media shape the flow of information to serve the interests of social classes; Littler, J. ( 2008), I Feel Your Pain: Cosmopolitan Charity and the Public Fashioning of the Celebrity Soul, explores how celebrity culture and the media act on class consciousness. G. (2006), Media Magic: Making Class Invisible, discusses how media contribute to class invisibility; Skeggs, B. (2004), Class, Self, Culture, explores how media representations Williams, R. (1977), Marxism and Literature, offers a theory of how media as a cultural practice constructs class relations. Identity formation and social stratification among young women are also future issues. Choukas-Bradley, S., and Prinstein, M. J. (2014) explore how peer influence and social media interact and how this affects young women's body image dissatisfaction and eating disorder behaviors Perloff, R. M. ( 2014) theoretically frames how social media affects young women's body image concerns and presents an agenda for future research. Fardouly, J., Diedrichs, P. C., Vartanian, L. R., and Halliwell, E. (2015), et al. examine how Facebook use affects young women's body image concerns and mood. Meier, E. P., and Gray, J. (2014) study how photographic activity on Facebook is associated with body image disturbance in adolescent girls; Rideout, V., Fox, S., Peebles, A., and Barr Taylor, C. ( 2016) et al. analyze how social media is linked to eating disorders in the digital age.

## 2.4 Regional Disparities Related to Social Class

Regional disparities related to social class have been studied in a wide variety of areas, including education, health, housing, and employment opportunities. This program provides case studies of social class disparities in a region and serves as a starting point for a better understanding of this issue. Each study was conducted in a specific national or regional context and explores how economic, social, and cultural factors create and maintain regional disparities. Sampson, R. J., Morenoff, J. D., and Gannon-Rowley, T. (2002) study the relationship between urban housing patterns and crime rates and assess how the social cohesion and organization of local communities affects class disparities; Putnam, R. D. (2000) analyzes the decline of social capital in the U.S. and explores how growing class inequality in local communities weakens social cohesion; Wilson, W. J. (1987), The Truly Disadvantaged: The Inner City, the Massey, D. S., and Denton, N. A. (1993), American Kasarda, J. D. (1995), "Industrial Restructuring," in Massey, D. S., and Denton, N. A. (eds.), Segregation and the Making of the Underclass, sheds light on the link between social class and regional disparities by showing how racial segregation in housing reinforces socioeconomic inequality. Industrial Restructuring and the Changing Location of Jobs, explores how industrial restructuring affects local economies and how this creates regional disparities that disadvantage lower-income groups. Hendren, N. and Katz, L. F. (2016), The Effects of Exposure to Better Neighborhoods on Children: New Evidence from the Moving to Opportunity Experiment studies how moving to better neighborhoods affects long-term outcomes for children through the Moving to Opportunity Experiment, evaluating the effects of class-based neighborhood disparities. Jargowsky, P. A. (1997), Poverty and Place: Ghettos, Barrios, and the American City, examines the problems concentrated in poor urban neighborhoods and how housing policies create and maintain class and regional disparities. Sharkey, P. (2013), Stuck in Place: Urban Neighborhoods and the End of Progress toward Racial Equality, tracks class and racial disparities in urban neighborhoods over time and examines the mechanisms by which these disparities mechanisms of how these disparities persist across generations.

## 2.5 The Relationship between Information Source Location Indicators and Social Class

The case study on the relationship between information source location indicators and social class provides a starting point for exploring how the introduction of information source location indicators can be useful in the context of social class, or how they can reveal information disparities between social classes. In particular, the focus will be on examining how information is distributed on the Internet and how it is accepted by people of different social classes; the study by Hargittai, E. (2002) examines how differences in Internet skills affect people's ability to access and process information and how people process and access information. This issue suggests a deep connection between social class and the adoption of information source location indicators. Blank, G., and Groselj, D. (2014) examine how various aspects of Internet use are linked to social class, arguing in particular that diversity of access to information sources is a central factor. Zhong, Z. J. (2013) study identifies how disparities in self-reports of digital skills affect the ways in which youth access and use information sources and considers how the introduction of source location indicators may be necessary. Van Deursen, A. J., and Van Dijk, J. A. (2014) show that the digital divide has shifted from simply a matter of access to differences in use and discuss in detail the impact of social class on access to information sources. Van Dijk, J. A. (2006) provides an overview of the achievements and limitations of digital divide research and discusses how the introduction of source location indicators could be beneficial; Robinson, L. et al. (2015) discusses how digital inequality is related to social class and discusses how tools such as source location index, and explores how tools like the index affect the ability of social class to process information. Research focusing on the relationship between social class and geographic conditions (often referred to as "geographic inequality" or "regional disparities") will also be important in how where people live affects education, employment, health, accessible services, etc. Kawachi, I and Berkman, L.F. ( Glaeser, E.L., Resseger, M., and Tobio, K. (2009), "Neighborhoods and Health," explores how neighborhoods affect residents' health and identifies links between social class and geographic differences in health. Wacquant, L. (2008), in Urban Outcasts: A Comparative Sociology of Advanced Marginality, provides an econometric analysis of inequality within cities and studies how urban geography affects social class. Wacquant, L. (2008), "Urban Outcasts: A Comparative Sociology of Advanced Marginality," analyzes social exclusion and class disparities in urban areas of developed countries from a comparative sociological perspective, exploring the impact of geographic segregation on social class; Putnam, R.D. (2015), "Our Kids: The American Dream in Crisis," discusses social mobility in the U.S. and the disparities between regions, and disparities between regions, and examines how access to education and economic opportunity varies geographically.

## 2.6 The Relationship Between Social Class and Educational Resources

Case studies on the relationship between social class and educational resources are an important factor with respect to how educational opportunities vary by social class and how this affects the future of individuals and groups.They provide both theoretical insights and empirical analysis of the allocation of educational resources and social class dynamics. Kozol, J. (1991), "Savage Inequalities: Children in America's Schools," provides a field study of the unequal distribution of educational resources in American schools, revealing educational disparities based on social class and race.Bourdieu, P., and Passeron, J.C. (1977), "Reproduction in Education, Society and Culture," analyzes how the educational system reproduces social inequalities and discusses the effects of social class on access to educational resources. Coleman, J.S.'s (1966) "Equality of Educational Opportunity," also known as the Coleman Report, examines the relationship between school resources, family environment, and student achievement, and how social class differences affect educational outcomes. Anyon, J. (1980), "Social Class and the Hidden Curriculum of Work," explores the "hidden curriculum" taught in schools belonging to different social classes, revealing the relationship between educational resources and social class. Ball, S.J. (2003), "Class Strategies and the Education Market: The Middle Classes and Social Advantage," studies how the middle class uses strategies to maximize educational opportunities for their children. Darling-Hammond, L. (2010), "The Flat World and Education: How America's Commitment to Equity Will Determine Our Future," examines how the middle class uses strategies to maximize educational opportunities for their children and how the education market affects social class. Reardon, S.F. (2011), "The Widening Academic Achievement Gap Between the Rich and the Poor: New Evidence," in Darling-Hammond, L. (2010), discusses how equity in education affects America's future and critically evaluates social class and the unequal distribution of educational resources. the Rich and the Poor: New Evidence and Possible Explanations" examines the widening academic achievement gap between children from economically affluent and poor families and elucidates the relationship between educational resources and social class.

## 2.7 Social Class and Digital Media Adoption

This case study on the relationship between social class and digital media explores how the digital divide, digital literacy, and online social interactions differ by social class.It provides a theoretical framework and empirical research on how social class influences the adoption and use of digital media and the resulting social consequences. To understand the role of social class in the digital age, these references provide a useful starting point.They deal with inequality in the access to and use of information technologies and are often key elements with respect to the social and economic context and its impact DiMaggio, P. et al. (2004) - DiMaggio et al.'s "Digital Inequality: From Unequal Access to Differentiated Use" explores various aspects of digital inequality, ranging from unequal Internet access to differential use. It analyzes how social class affects access to and use of digital media.Hargittai, E. (2002) - Hargittai's Second-Level Digital Divide: Differences in People's Online Skills" analyzes the impact of social class on digital literacy and takes a closer look at inequalities in online skills Norris, P. (2001) - Norris's "Digital Divide: Civic Engagement, Information Poverty, and the Internet Worldwide examines how the Internet affects civic engagement and information inequality worldwide, and identifies the link between digital media and social class Selwyn, N. (2004) - Selwyn's Reconsidering Political and Popular Understandings of the Digital Divide," questions popular and policy understandings of the digital divide and explores the connection between social class and digital access and use. van Dijk, J.A.G.M. (2005) - van Dijk's The Deepening Divide: Inequality in the Information Society analyzes the growing inequality in the information society and examines how the introduction of digital media affects social class inWarschauer, M. (2003) - Warschauer's Technology and Social Inclusion: Rethinking the Digital Divide explores how technology can contribute to social inclusion, It explores how the digital divide can be overcome and offers a new perspective on the relationship between social class and digital media.

## 2.8 Digital and Regional/National Disparities

"Digital Inequalities and Why They Matter" by Robinson, L., et al. (2015) discusses in detail the ways in which access to digital technologies is not provided equally to all people and how this affects society. "The Digital Divide: The Internet and Social Inequality in International Perspective" by Ragnedda, M., and Muschert, G. W. (2013), provides an international perspective on how the InternetThe authors discuss how the Internet contributes to social inequality from an international perspective."It's Complicated: The Social Lives of Networked Teens" by boyd, d. (2014) provides a detailed case study of how young people use digital media and how social class influences their usage and how social class influences their usage."Distinct Skill Pathways to Digital Engagement" by Helsper, E. J., and Eynon, R. (2013) examines how engagement with digital technologies varies by individual skill and social class. "Participatory Culture in a Networked Era: A Conversation on Youth, Learning, Commerce, and Politics" by Jenkins, H., Ito, M., and boyd, d. (2016), examines the digital participation culture and how it relates to social class and education."Gradations in Digital Inclusion: Children, Young People and the Digital Divide" by Livingstone, S., and Helsper, E. J. (2007) focuses on how youth and children Bourdieu, E. J. (2007), "Gradations in Digital Inclusion: Children, Young People and the Digital Divide."The Forms of Capital" by Bourdieu, P. (1986) provides the concepts of economic, social, and cultural capital as a basis for understanding how these affect the formation of social classes and the use of digital media in the digital age."Digital Skills: Unlocking the Information Society" by Van Deursen, A. J., and Van Dijk, J. A. (2014), examines how digital skills vary across social classes and how they DiMaggio, P., and DiMaggio, P. (2014), "Digital Skills: Unlocking the Information Society.

DiMaggio, P., and Hargittai, E. (2001) analyze the transition from "digital divide" to "digital inequality" that occurs as the Internet becomes increasingly popular and consider various factors, including regional differences Norris, P. (2001) studies how Internet Servon, L. J. (2002) explores how local communities and public policy bridge the digital divide and suggests ways to enhance digital literacy in the context of regional disparities. Ragnedda, M., and Muschert, G. W. (2013) work explores the relationship between the Internet and social inequality from an international perspective and includes a case study on digital literacy gaps across regions. Graham, M. (2011) focuses on the spatial aspects of digital disparities and discusses the impact of geographic location on digital literacy. The report by Zickuhr, K., and Smith, A. (2012) provides a detailed analysis of regional differences in Internet use within the U.S. The study by Bertot, J. C., Jaeger, P. T., and McClure, C. R. (2008) provides an in-depth look at how eAttewell, P. (2001) identifies early and late gaps in digital literacy and assesses how these relate to regional differences.The study by Vicente, M. R., and López, A. J. (2010) analyzes how regional disparities affect Internet use through a Spanish case study.

## 2.9 Media environment and class disparity

Research on media environment disparities and class disparities explores how media access, use, and content vary by socio-economic status. The literature on this ranges from a focus on media literacy, fair access to information, and its correlation with social class to consideration of how particular media reflect and shape class consciousness and stereotypes. There is a wide range of We provide deep insight and detailed research that helps us understand how media is experienced differently by class and how socio-economic status shapes disparities in media access and use. Furthermore, it is important to closely examine how disparities in the media environment affect the social position of individuals and the class structure of society as a whole. Research on equitable access to information and class differences is also important in examining how information and communication technologies (ICTs) create different access and usage patterns between high and low socio-economic groups. Couldry, N. (2010), Why Voice Matters: Culture and Politics After Neoliberalism, highlights the importance of voice in culture and politics after neoliberalism, and argues that the media widens the gap in class communication. We are analyzing the situation. Misunderstanding the Internet, by Curran, J., Fenton, N., and Freedman, D. (2012), critically examines how the democratizing promise of the Internet is misunderstood, and We analyze situations in which economic groups are excluded from the information society. DiMaggio, P., Hargittai, E., Celeste, C., and Shafer, S. (2004), "Digital Inequality: From Unequal Access to Differentiated Use," discusses not only the unequal access to the Internet, but also the diversity of use. I am also researching gender and exploring how these differences influence class disparities. Bourdieu, P. (1998)'s On Television and Journalism critically analyzes the role of television and journalism, examining how the media reflects and solidifies class structures. Hesmondhalgh, D. (2007)'s The Cultural Industries explores how cultural industries contribute to the formation of social classes and how media content differs by class. Murdock, G., and Golding, P. (2005)'s Culture, Communications, and Political Economy examines the interplay between culture, communication, and political economy and how this relates to the media environment and class divide. We are analyzing whether there are any. Tichenor, P. J., Donohue, G. A., and Olien, C. N. (1970), Mass Media Flow and Differential Growth in Knowledge, studies the influence of media on the distribution of knowledge and how this differs between classes. and proposes the "knowledge gap hypothesis". Garnham, N. (2000)'s Emancipation, the Media, and Modernity: Arguments About the Media and Social Theory discusses the relationship between media and modern society and explores how media is embedded in class structures. Research that focuses on the negative effects of media related to social class examines how the media contributes to reinforcing class stereotypes and justifying or ignoring economic inequalities. The research that is being done is outstanding. There are important insights into the role of the media in relation to economic and social class disparities, revealing the various ways in which the media contribute to the entrenchment of class structures and the generation of stereotypes. McChesney, R. W. (2013), Digital Disconnect: How Capitalism is Turning the Internet Against Democracy, discusses how the relationship between the Internet and capitalism is affecting democracy, especially the negative impact on economically disadvantaged classes. It points out the impact. Couldry, N. (2000)'s The Place of Media Power: Pilgrims and Witnesses of the Media Age provides a theoretical analysis of how the media shapes social power and reproduces class inequality. . The chapter on the relationship between the Internet and social class in Misunderstanding the Internet by Curran, J., Fenton, N., and Freedman, D. (2016) explores how the Internet reinforces existing social class inequalities. We are digging deep into what is going on. Davis, A. (2019)'s Professional Identity Crisis: Race, Class, Gender, and Success at Professional Schools explores the dynamics of success around race, class, and gender in professional schools, and explores the media's focus on these issues. The focus is on how you handle it. Tyler, I. (2013)'s Revolting Subjects: Social Abjection and Resistance in Neoliberal Britain examines social exclusion and resistance in neoliberal Britain, exploring how the media shapes and reproduces class distinctions. We are discussing what is going on.

Skeggs, B., and Wood, H. (2012), "Reacting to Reality Television: Performance, Audience and Value," examines how reality television constructs class consciousness in viewers and creates stereotypes of particular social classes. We are analyzing how it is being strengthened. Mantsios, G. (2006)'s "Media Magic: Making Class Invisible" critically examines how the media obscures class issues and explains the mechanisms that obscure class disparities.

Elite Discourse and Racism by Dijk, T. A. van (1993) explores how elite-led discourses influenced issues of race and class, and how the media involved them in the process. This is what Dijk, T. A. van et al. (1993) considered.

## 2.10 The Knowledge Gap Hypothesis

The Knowledge Gap Hypothesis is the theory that when the media provides information, people of higher socioeconomic status obtain information at a faster rate, resulting in greater knowledge inequality. It lays the foundation for understanding how the media affect political knowledge and participation. Each study explores the relationship between the knowledge gap hypothesis and political participation from different perspectives, with a particular focus on how media usage and political communication patterns affect an individual's political participation. These references may also serve as a starting point for a detailed understanding of the interplay between the knowledge gap and political participation. Some studies on the knowledge gap hypothesis and Socio-Economic Status (SES) explore how the distribution of information and access to education create inequalities among specific socioeconomic groups. Tichenor, P. J., Donohue, G. A., and Olien, C. N. (1970), Mass Media Flow and Differential Growth in Knowledge, points out that information distributed by the media has a different effect on higher and lower educational attainment, producing knowledge inequality, This was the first study to propose the knowledge gap hypothesis. This is the first study to propose the knowledge gap hypothesis. Kwak, N. (1999) reassesses how education and motivation shape patterns of media use and thus widen the knowledge gap in Viswanath, K. and Finnegan, J. R. (1996) in their paper, The Knowledge Gap Hypothesis: Twenty-Five Years Later, reviews the last 25 years of research on the knowledge gap hypothesis and discusses the effects of media and social status on knowledge acquisition. The Knowledge Gap Hypothesis is the theory that when the media provide information, people of higher socioeconomic status acquire information at a faster rate, resulting in greater knowledge inequality. It lays the foundation for understanding how the media affect political knowledge and participation. Each study explores the relationship between the knowledge gap hypothesis and political participation from different perspectives, with a particular focus on how media usage and political communication patterns affect an individual's political participation. These references may also serve as a starting point for a detailed understanding of the interplay between the knowledge gap and political participation. Some studies on the knowledge gap hypothesis and Socio-Economic Status (SES) explore how the distribution of information and access to education create inequalities among specific socioeconomic groups.

Tichenor, P. J., Donohue, G. A., and Olien, C. N. (1970), Mass Media Flow and Differential Growth in Knowledge, points out that information distributed by the media has different effects on higher and lower educational attainment, This was the first study to propose the knowledge gap hypothesis. This is the first study to propose the knowledge gap hypothesis. Kwak, N. (1999) reassesses how education and motivation shape patterns of media use and thus widen the knowledge gap in Viswanath, K. and Finnegan, J. R. (1996) in their paper, The Knowledge Gap Hypothesis: Twenty-Five Years Later, reviews the last 25 years of research on the knowledge gap hypothesis and discusses the effects of media and social status on knowledge acquisition. In The Knowledge Gap: An Analytical Review of Media Effects, Gaziano, C. (1983) provides a comprehensive review of media effects and analyzes how differences in information processing across classes shape the knowledge gap Eveland, W. P., and Scheufele, D. A. (2000) explore how news media use is related to knowledge acquisition and political participation. Hwang, Y., and Jeong, S. H. (2009), through "Revisiting the Knowledge Gap Hypothesis: A Meta-Analysis of Thirty-Five Years of Research," examine the effects of education and media access on knowledge gaps. Eveland, W. P. Jr. and Hively, M. H. (2009) examine how the frequency of political debate, network size, and diversity of debate affect political knowledge and participation. In Fraile, M. (2011), Widening or Reducing the Knowledge Gap? Testing the Media Effects on Political Knowledge in Spain (2004-2006), examines how the media have affected the political knowledge of Spanish citizens. Iyengar, S., and Hahn, K. S. (2009) in their article "Red Media, Blue Media: Evidence of Ideological Selectivity in Media Use" explores the relationship between Internet use and civic participation, specifically how selective media In particular, it examines how selective media use affects political knowledge and participation.

Gil de Zúñiga, H., Jung, N., and Valenzuela, S. (2012) explore how the use of news via social media impacts social capital, civic engagement, and political participation. Scheufele, D. A., and Nisbet, M. C. (2002) study, Being a Citizen Online: New Opportunities and Dead Ends, examines whether online civic engagement offers new possibilities or limits Tichenor, P. J., Donohue, G. A., and Olien, C. N. (1970) published Mass Media Flow and Differential Growth in Knowledge, a foundational study showing that information distributed by the media has different effects on higher and lower educated populations, creating knowledge gaps. Differential Growth in Knowledge. Viswanath, K., and Finnegan, J. R. Jr. (1996) in their paper, The Knowledge Gap Hypothesis: Twenty-Five Years Later, review 25 years of research on the knowledge gap hypothesis and the social implications for understanding. Kwak, N. (1999) in his study, "Revisiting the Knowledge Gap Hypothesis: Education, Motivation, and Media Use," reassesses how education, motivation, and media use affect the knowledge gap Eveland, W. P., and Eveland, W. P. Eveland, W. P. Jr. and Scheufele, D. A. (2000), "Connecting News Media Use with Gaps in Knowledge and Participation," analyzes the relationship between news media use and knowledge and political participation gaps. The article analyzes the relationship between news media use, knowledge, and gaps in political participation. Holbert, R. L., Garrett, R. K., and Gleason, L. S. (2010), "A New Era of Minimal Effects? A Response to Bennett and Iyengar," assesses the impact of the new media environment on the knowledge gap hypothesis and examines the relationship between media effects and the knowledge gap. Hindman, D. B. (2009), "Mass Media Flow and Differential Access to Political Information: The Knowledge Gap and the Digital Divide," discusses the impact of new media environments on the knowledge gap hypothesis. Hindman, D. B. (2009), Mass Media Flow and Differential Access to Political Information: The Knowledge Gap and the Digital Divide, examines how the imbalance between information flow through the media and access to political information affects the knowledge gap and the digital divide. In Fraile, M. (2013), in his article "Do

Information-Rich Contexts Reduce Knowledge Inequalities? The Contextual Determinants of Political Knowledge in Europe, The Contextual Determinants of Political Knowledge in Europe, the authors focus on the contextual determinants of political knowledge in Europe and consider the potential for information-rich environments to reduce knowledge inequalities.

Viswanath, K., and Finnegan, J. R. Jr. (1996), The Knowledge Gap Hypothesis: Twenty-Five Years Later, reevaluates the relationship between SES (socioeconomic status) and knowledge acquisition and reexamines the knowledge gap hypothesis 25 years later. Kwak, N. (1999), Revisiting the Knowledge Gap Hypothesis: Education, Motivation, and Media Use, examines how education, motivation, and media use affect the knowledge gap and reexamines the knowledge gap hypothesis 25 years later. Hargittai, E. (2002), Second-Level Digital Divide: Differences in People's Online Skills, examines how education, motivation, and media use affect the knowledge gap and focuses on the relationship with socioeconomic status (SES). DiMaggio, P., and Hargittai, E. (2001), From the 'Digital Divide' to 'Digital Inequality': Studying Internet Use as Penetration Increases," shifts the perspective from digital divide to digital inequality and identifies SES-based inequalities as the Internet becomes increasingly popular. 2002), The Internet and Knowledge Gaps: A Theoretical and Empirical Investigation, explores theoretically and empirically the relationship between the Internet and the knowledge gap, examining the impact of SES on the ability to acquire information online. Gaziano, C. (1983), The Knowledge Gap: An Analytical Review of Media Effects, advances our understanding of the relationship between SES and knowledge acquisition through an analytical review of media effects. Marr, M. (2005), Digital Divide: Civic Engagement, Information Poverty, and the Internet Worldwide, explores Internet penetration and information poverty from a global perspective and discusses how SES Witte, J. C., and Mannon, S. E. (2010), The Internet and Social Inequalities, examines the relationship between the Internet and social inequalities and how SES affects access to online resources.

## 2.11 Racial Segregation

Research on racial segregation ranges from residential segregation based on race, to school segregation, to inequality in access to employment and other social resources. Wilson, W. J. (1987), The Truly Disadvantaged: The Inner City, the Underclass, and Public Policy, focuses on segregation and socioeconomic disadvantage in inner-city poor neighborhoods and offers policymakers Krysan, M., and Crowder, K. (2017), Cycle of Segregation: Social Processes and Residential Stratification, explores how housing choice can reinforce existing segregation patterns and provides deep insights into the social processes of residential location by investigating how residential choice reinforces existing patterns of segregation. Rothstein, R. (2017), The Color of Law: A Forgotten History of How Our Government Segregated America, provides a historical Pager, D. (2003), The Mark of a Criminal Record, studies the impact of criminal records on racial segregation in the job market and explores the broader labor market consequences of the criminal justice system; Logan, J. R., and Stults, B. J. (2011), The Persistence of Segregation in the Metropolis: New Findings from the 2010 Census, draws on 2010 Census data to Sharkey, P. (2013), Stuck in Place: Urban Neighborhoods and the End of Progress toward Racial Equality, examines how racial inequality in low-income urban neighborhoods has persisted for generations. Sharkey, P. (2013), explores how racial inequality persists over generations in low-income urban neighborhoods and has detailed barriers to social mobility.

## 2.12 Social Class of Filter Bubble

Research focusing on the relationship between social class and filter bubbles is important to note because of the risk that personalization of information on the Internet reinforces social hierarchies and may create information disconnects between different classes. It provides a theoretical and empirical basis for how filter bubbles affect different social classes differently and the potential risks they offer for further social fragmentation. Researchers and policy makers could use these findings to explore ways to mitigate the information disconnects that the Internet creates. Pariser, E. (2011)Pariser's book, The Filter Bubble: What the Internet is Hiding from You, introduces the concept of the filter bubble and discusses how individual online searches limit an individual's worldview and the social class-based access to Sunstein, C. R.(2017)Sunstein's Republic: Divided Democracy in the Age of Social Media discusses how social media creates social class-based echo chambers and analyzes what this means for democracy. Bakshy, E., Messing, S., and Adamic, L. A. (2015) - Bakshy et al.'s research paper, Exposure to ideologically diverse news and opinion on Facebook, examines how Facebook users are exposure to diverse news and opinions and how socioeconomic status affects information exposure; Zuiderveen Borgesius, F. J., et al. (2016) - Zuiderveen Borgesius et al. in their study "Should we worry about filter bubbles?" explores how online information filtering affects social hierarchies; Liao, Q. V., and Fu, W. T. (2013) - Liao and Fu's study "Can You Hear Me Now? the Echo Chamber Effect by Source Position Indicators," investigates the effects on social class of the introduction of source position indicators to mitigate the echo chamber effect. hargittai, E., and Hinnant, A. (2008) - Hargittai, E., and Hinnant, A. (2008). Hargittai and Hinnant's study, Digital Inequality: Differences in Young Adults' Use of the Internet, examines digital disparities in young adults' Internet use and the socioeconomic

The study analyzes the impact of socioeconomic background on online information acquisition. Mitchell, A., et al. (2014) - Mitchell et al.'s study, Political Polarization and Media Habits, explores political polarization and media usage habits and examines how different social classes shape information bubbles.

# 3. Disparity related to the Five mechanisms of maintaining inequality in society

Disparity issues related to the five mechanisms of maintaining inequality in society (Elitism, Exclusionism, Exclusion, Scheduling, and Shaming) In understanding how inequality is maintained within a social system, the study of the five mechanisms of maintenance of inequality in society involves the study of how power and status are established, maintained, and transmitted through social, educational, and economic processes. These studies can also examine how these mechanisms affect individual identity, social relationships, and overall social cohesion.

## 3.1 Disparity Issues Related to Elitism

Case studies on disparity issues related to elitism provide a perspective from which we can analyze how elites maintain privilege and create social and economic disparities in various areas, including education, corporate governance, social networks, and policy-making processes. Exclusivism refers to social practices in which certain social groups or communities intentionally exclude other groups and thus limit their access to economic and social resources. Exclusivism is known to create social divisions and reinforce social inequalities. Khan, S. R. (2011) focused on the elite formation process in prestigious American boarding schools and conducted a study examining the reproduction of elitism through education. Young-Bruehl, E. (1996) sheds light on the mechanisms by which certain social groups exclude others through the analysis of prejudice. Wacquant, L. (2008) details the mechanisms and effects of economic and social exclusion in urban environments. Bourdieu, P., and Passeron, J. C. (1977) elucidates the ways in which educational systems reproduce existing social structures as "scheduled harmony." Sennett, R., and Cobb, J. (1972) et al. explored the impact of class-based social status and its attendant shaming on individual self-esteem and identity. Mills, C. W. (1956) et al. discussed how economic, political, and military elites in American society concentrate power and influence society, Pakulski, J., and Waters, M. (1996) et al. examined how globalization affects the existing class system and how elites maintain their position. Hartmann, M. (2007) analyzed the role of elites in contemporary society and how they use economic and cultural capital to maintain their position. It describes how elites within large corporations exercise power and influence social and economic inequality through corporate governance. Useem, M. (1984) et al. explained how elites within large firms exercise power and influence social and economic inequality through corporate governance. Bourdieu, P. (1984) et al. explored how cultural preferences shape social class and how elites maintain their status through cultural superiority. It provides deep insights into how elitism functions in diverse aspects of society. The approach revealed how the behavior of elites, their values, and their use of social and economic resources contribute to the continuation of widespread social inequality.

## 3.2 Exclusivism

To help understand how exclusivism shapes and sustains economic and social inequalities, practices of exclusivism in various social domains such as education, housing, employment, and public policy are surveyed. Exclusivism is a behavior or attitude in which one group intentionally marginalizes or excludes another group. This is often based on attributes such as race, ethnicity, religion, gender, sexual orientation, and disability status. Wacquant, L. (1998) in "Urban Outcasts: A Comparative Sociology of Advanced Marginality." conducted an in-depth study of the relationship between marginalized groups and exclusive social structures in urban settings. Massey, D. S., and Denton, N. A. (1993) examine how racial segregation in the U.S. housing market deepens economic disparities. Young, I. M. (2000) analyzed the impact of exclusivism on democratic ideals and proposed a theoretical framework to promote a more inclusive public life. Kymlicka, W. (1995) discusses minority rights and liberal responses to exclusivism in multicultural societies. Charles, C. Z. (2003) mentioned how housing segregation affects the interaction between social classes and how it maintains disparities. Anderson, E. (2010) argued why social integration is essential to social justice and examined the negative effects of social division through exclusionary practices.

## 3.3 Scheduled Harmony

Literature focusing on disparity issues related to scheduled harmony includes the following studies that explore scheduled harmony as a false representation of structural inequality and social harmony, and the reproduction of inequality by social and economic institutions. Bourdieu, P., and Passeron, J. C. (1977) provides a classic analysis of how educational systems reproduce social inequality. Marcuse, H. (1964) in his book The Unitary Man: An Ideological Study of Advanced Industrial Societies analyzes how advanced industrial societies create unidimensional thought patterns and pseudo-needs and how these reinforce existing social imbalances and acceptance. Milanovic, B. (2016), in his book Global Inequality: a New Approach in the Age of Globalization, ex-plores the causes of inequality and its variation from a global perspective, offering a view of global equilibrium. Stiglitz, J. E. (2012), in his book, The Price of Inequality: How Today's Divided Societies Put Our Future at Risk, analyzes how social equilibrium actually exacerbates economic inequality and proposes a path toward more equal economic development. Wilkinson, R., and Pickett, K. (2009) in their book The Spirit Level: Why More Equal Societies Are Almost Always Better reveals that economic inequality has multifaceted negative effects in society and raises questions in terms of equality and balance In. Young, M. D.'s (1958) work, The Rise of Meritocracy, develops an imaginative perspective on how merit-based societies form new hierarchies and the resulting social balance. These advance our understanding of the reproduction of disparities in areas such as education, economics, and social policy and how they support social scheduled harmony. Issues of social scheduling are often discussed at the intersection of various social identities, such as class, race, and gender.

### 3.4 Cases of Humiliation

Literature exploring how the experience of humiliation and indignity is linked to social disparities and inequalities is often found in the fields of social psychology, sociology, and anthropology. These are studied in terms of the impact of humiliation on personal dignity and social identity, and how it shapes an individual's position within the social hierarchy. Sennett, R., and Cobb, J. (1972) in their book, The Hidden Wounds of Class, explore how individuals belonging to the lower social strata experience humiliation in their daily lives and how it affects their self-esteem.Smith, A.'s (2001) work, The Theory of Moral Sentiments, is an economics classic in which Adam Smith discusses the human instinct to seek the approval of others and regarding social humiliation.Walker, R. (2014)'s work, The Shame of Poverty, analyzes how poverty is associated with humiliation from a global perspective.Goffman, E. (1963) in his book Stigma: The Social Psychology of Stigma explores how social stigma and humiliation affect an individual's self-perception from a social psychology perspective. Hochschild, A. R. (1983), in his book, The Managed Heart: The Commodification of Human Emotion, discusses through case studies how emotional labor can create humiliation in the service industry.Wilkinson, R. G. (2005), The Impact of Inequality: How to Make a Sick Society Healthy, examines how economic inequality amplifies humiliation in society from an epidemiological perspective. Lindner, E. G. (2006) in his book, Creating Enemies: Humiliation and International Conflict, examines how humiliation in international relations intensifies conflict. It studies the effects of humiliation on the psychological health of individuals, social relationships, and even relations between social classes, often associated with disparity and the experience of living in communities of low socioeconomic status.

### 3.5 Cases of Meritocracy

Sociologists and political theorists have conducted a variety of analyses of social hierarchies and scheduling as they relate to meritocracy. They have identified mechanisms by which meritocracy, while promising equality and efficiency in theory, in practice creates conditions that favor certain classes and groups and reinforce existing social hierarchies. Young, M. (1958), in his work "The Rise of Meritocracy," critically satirizes how meritocracy contributes to the creation of a new elite class and the maintenance of social homogeneity and inequality. Meritocracy and Economic Inequality, edited by Arrow, K. J., Bowles, S., and Durlauf, S. N. (2000), explores how economic imbalances are created and maintained under meritocratic regimes.Littler, J. (2017), in his book, Against Meritocracy: the Myth of Culture, Power, and Mobility, offers a critical perspective on the myth of meritocracy that takes root in contemporary culture and the social and economic imbalances it creates.McNamee, S. J., and Miller Jr, R. K. (2004), The Myth of Meritocracy, focuses on the gap between the ideals and realities of meritocracy in American society and how this regime fixes hierarchy.Khan, S. (2011), in his book Privilege: the Making of a Youth Elite at St. Paul's School, shows through case studies how elite education is involved in the reproduction of social hierarchy.Rivera, L. A. (2015), in his book Pedigree: How Elite Students Get Elite Jobs, analyzes how meritocracy in the labor market actually works and why graduates of certain elite universities are more likely to find jobs in high-ranking professions.Sandel, M. J.'s (2020) book, The Tyranny of Merit: Where Has the Common Good Gone? discusses the contemporary social challenge that meritocracy poses to the common good. It provides a starting point for further discussion of how meritocracy creates social scheduling and creates new hierarchies. Each author describes how meritocracy affects social and economic inequality.

### 3.6 Cases of Scheduled Harmony

The sociological theory of scheduled harmony (functionalism) is based on the idea that each element of society exists for the stability and function of the whole and that each part contributes to the whole.However, criticism exists that this idea favors certain classes or groups. Much of the literature cited here is research that examines how the theory of scheduled harmony actually benefits certain social classes and marginalizes others. Parsons, T. (1951), in his work "Social Systems," discusses how each component of society functions as a whole and contributes to one another, and this theory was later developed by scholars and criticized as providing biased benefits to certain classes.Merton, R. K. (1968), in his work Social Theory and Social Structure, ex-tends the concept of scheduled harmony theory, noting that social structure can sometimes be dysfunctional and exploring its effects.Dahrendorf, R. (1959), in his work "Class and Class Struggle in Industrial Society," provides valuable insights into theories of social change and conflict by critiquing class conflict and the concentration of interests in particular classes, which scheduled harmony theory overlooks.Wright Mills, C. (1956), in his work "Power Elite," argues that the scheduled harmony of society actually helps the elite class maintain power and control over other classes.Bourdieu, P. (1984), in his work "Distinction: A Social Critique of Taste," uses the concept of cultural capital to elucidate the mechanisms by which certain classes maintain their social status and exclude others.Collins, R. (1979), in his work "Credential Society: A Historical Sociology of Education and Stratification," discusses how educational credentials function as a means of fixing social class and benefiting particular groups. It provides an important perspective from which to analyze the interactions between social structures and individuals, particularly the inequalities that schedules bring about in terms of power, status, and class.Each author reveals how social functionalism supports social structures that appear harmonious but in fact continue to produce inequality.

## 4. Opinions of the Six Social Clusters

When looking at the distribution of the six types of opinions, the hypothesis calculations for flexible social conditions are first addressed in this paper. We dare to assign the following parameter settings with random numbers with values ranging from 0.1 to 0.99. First, the value of $A$ is the value of trust in one's own opinion, and the rule is to assign larger values in the order $a > f$. As for $B$, we want to assign it a value between 0.1 and 0.99 completely at random, since it is an effect of the external efficacy. $C$ should be assigned a high score at random if $A$ has high confidence in his/her opinion and $B$ has a low value.

To simulate the dynamics of opinions across the six social classes ("$a$" through "$f$") and the flexible social conditions specified, we adopt a probabilistic approach to parameter assignment. This approach takes into account the specified rules for assigning values to parameters $A$, $B$, and $C$.

The rules for assigning values are understood and potentially implemented programmatically in the following ways

Parameter $A$ (confidence value of self-opinion):.

Values for $A$ are assigned in descending order such that $a > f$. This means that class '$a$' must have the highest value of $A$, class '$f$' must have the lowest value, and the other classes are in descending order from '$a$' to '$f$'. Parameter $B$ (external/media influence):.

The value of $B$ is assigned completely randomly within the range of 0.1 to 0.99. Parameter $C$ (Influence of Invariant Fields): The value of C is assigned completely randomly

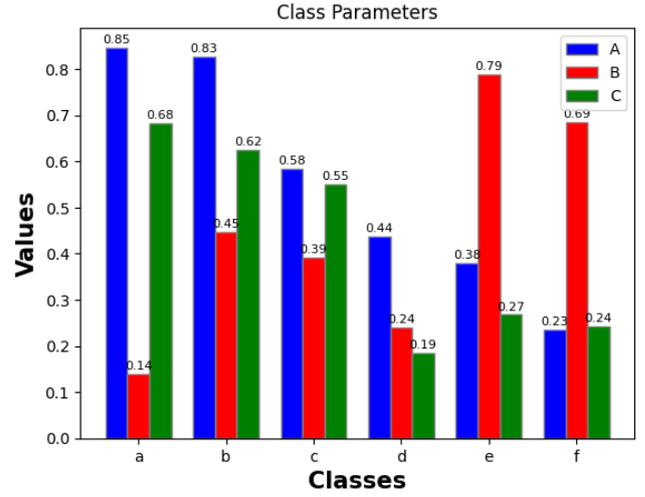

Fig. 1: Class Parameters

within the range of 0.1 to 0.99.

Indicates the conservative nature of the opinion. If $A$ is high and $B$ is low, a higher value of $C$ is assigned at random, implying a more conservative class. The settings are based on these rules.

Given a set of classes $a, b, c, d, e, f$, we assign the parameter values for $A$, $B$, and $C$ as follows:

The value for $A$ is assigned in descending order from a randomly generated uniform distribution between 0.1 and 0.99.

The value for $B$ is assigned randomly from a uniform distribution between 0.1 and 0.99 for each class independently.

The value for $C$ is conditionally assigned based on the values of $A$ and $B$. Specifically,

$$\text{If } A > 0.5 \text{ and } B < 0.5:$$
$$C \sim \text{Uniform}(0.5, 0.99),$$
$$\text{otherwise:}$$
$$C \sim \text{Uniform}(0.1, 0.49).$$

### 4.1 Opinion Dynamics Simulation Considering Class Sizes

In the simulation of opinion dynamics, we consider a population divided into different classes. The interaction between individuals within and across these classes is influenced by various factors, including the total opinions in each class and the sizes of the classes.

(1) **Total Opinions in a Class**:
$$\text{TotalOpinions}_{\text{class}} = \sum_{i \in \text{class}} \text{opinions}[i] \qquad (1)$$

(2) **Interaction Coefficient between Classes**:

$$D_{\text{class1,class2}} = \frac{\text{TotalOpinions}_{\text{class2}} \times A_{\text{class1}}}{\text{class\_sizes}[\text{class2}]} \quad (2)$$

This equation represents the influence exerted on individuals in *class1* by the entire *class2*.

(3) **Update Rule for Opinions**:

$$\begin{aligned}
\text{opinions}_{\text{new}}[i] =\ & \text{opinions}_{\text{old}}[i] \\
& + \sum_{j \neq \text{class}} D_{\text{class},j} \cdot \text{size}[j] \\
& - C_{\text{class}} \cdot \text{opinions}_{\text{old}}[i] \\
& + B_{\text{class}}
\end{aligned} \quad (3)$$

The model described above provides a foundation for simulating opinion dynamics in a population segmented into distinct classes. By taking into account the sizes of each class and the interactions between them, we can achieve a more nuanced and accurate representation of how opinions evolve over time.

# 5. Trust-Distrust Model, Opinion Distribution

It simulates opinion dynamics within a group with multiple different classes. Each class has a different confidence score based on different initial opinions, external influences, and opinion constancy. Opinions fluctuate over time based on interactions with other individuals, external influences, and the individual's internal credibility.

This models the opinion dynamics in different classes (a through f) of individuals over a period of time. These opinions are influenced by internal and external factors as well as interactions between individuals.

## 5.1 Parameters and Variables

**class_sizes**: A dictionary specifying the number of individuals in each class.

**class_parameters**: A dictionary that sets the trust score (A), trust in external influence (B), and trust in invariance of opinion (C) for each class.

**D**: A matrix representing the trustworthiness between individuals, initialized randomly.

**class_initial_opinions**: A dictionary that sets the range of initial opinions for each class.

**opinions**: An array storing the initial opinion of each individual based on their class.

**class_indices**: An array mapping each individual to their respective class.

## 5.2 Opinion Dynamics Simulation

The function *simulate_opinion_dynamics* simulates the evolution of opinions over a given number of time steps. It employs the following logic:

$$\begin{aligned}
\text{opinion}_{i,t+1} =\ & \text{opinion}_{i,t} \\
& + \sum_{j \neq i} D_{i,j} \times \text{opinion}_{j,t} \times \text{trust\_to\_others} \\
& + \text{external\_influence} \\
& - \text{invariance} \times \text{opinion}_{i,t}
\end{aligned} \quad (4)$$

where $D_{i,j}$ represents the trustworthiness between individual $i$ and individual $j$.

Opinion trajectories are plotted for each class at certain intervals. These trajectories represent the mean opinion value over time. Additionally, histograms are used to showcase the distribution of opinions at the end of the simulation.

## 5.3 Opinion Transition Detection

The code contains a function *detect_transition* which identifies transitions in the collective opinion of each class. Transitions are categorized as:

> Neutral to Positive
> Positive to Neutral
> Positive to Negative
> Negative to Positive
> Neutral to Negative
> Negative to Neutral

Bar charts are utilized to visualize the transitions in each class over certain intervals.

This code offers a detailed simulation of opinion dynamics in different classes. The opinions evolve based on internal predispositions, external influences, and interactions with other individuals, and the transitions in these opinions can be visualized using the provided visualization tools.

## 5.4 Function for Clustering Opinions

The function `cluster_opinions` receives an individual's opinion as input and categorizes it into one of three categories: Negative, Neutral, or Positive.

$$\text{cluster\_opinions}(o) = \begin{cases} \text{Negative} & \text{if } o < \theta_1 \\ \text{Neutral} & \text{if } \theta_1 \leq o < \theta_2 \\ \text{Positive} & \text{if } o \geq \theta_2 \end{cases} \quad (5)$$

Where:

> $o$ represents the individual's opinion.
>
> $\theta_1$ and $\theta_2$ are the thresholds for categorizing the opinions.

### 5.5 Graph for Opinion Clusters Count

The function `plot_opinion_clusters` plots the distribution of opinions at each time step. For each class, it counts the number of opinions falling into each category.

$$\text{class\_cluster\_counts}[c] = \text{Counter}(\text{clusters}[\text{class\_indices} == i]) \quad (6)$$

Where:

> $c$ denotes the class ID.
>
> clusters is a list containing the category of each opinion.
>
> class_indices is a list of indices corresponding to each class.

### 5.6 Graph for Inter-opinion Trust Scores

The function `compute_trust_scores` calculates the trust scores for each individual.

$$\text{trust\_scores}[i] = \sum_{j=1}^{N} D_{ij} \quad (7)$$

Where:

> $D$ is the trust matrix.
>
> $N$ represents the total number of individuals.
>
> $i$ and $j$ are indices denoting individual opinions.

### 5.7 Function to Detect Opinion Transitions

In this section, we explain a function designed to detect transitions in group opinions at different time points, and the process to count and visualize these transitions.

### 5.8 Transition Detection Function: Detect_Transition

This function takes two consecutive opinion values as its arguments and returns the type of opinion transition.

$$\text{Detect\_Trans}(\text{prev}, \text{current}) :$$

$$\begin{cases} \text{"Positive to Neutral"} & \text{if prev} > 0 \text{ and current} \leq 0, \\ \text{"Neutral to Positive"} & \text{if prev} \leq 0 \text{ and current} > 0, \\ \text{"Positive to Negative"} & \text{if prev} > 0 \text{ and current} < -50, \\ \text{"Negative to Positive"} & \text{if prev} < -50 \text{ and current} > 0, \\ \text{"Neutral to Negative"} & \text{if prev} \geq -50 \text{ and current} < -50, \\ \text{"Negative to Neutral"} & \text{if prev} < -50 \text{ and current} \geq -50, \\ \text{None} & \text{otherwise.} \end{cases}$$

### Extension of Opinion Dynamics Simulation

In this section, we break down an extended function for simulating opinion dynamics and its associated visualization.

### Extended Simulation Function: `Simulate_Opinion_Dynamics_Extended`

The function simulates how individual opinions evolve over time. It takes into account:

> The current opinions of individuals.
>
> Influence from others.
>
> External influences.
>
> A degree of invariance (resistance to change).

> Input: opinions, D, class_parameters_list, class_indices, time_steps
>
> Output: opinion_history, influence_history, invariance_history

**Algorithm:**

(a) Initialize histories of opinions, influences, and invariance values.

(b) For each time step:

> Calculate the influence on each individual from all others, considering trust and external influence. Store this in `Influence_History`.
>
> Calculate invariance for each individual based on their own opinion. Store this in `Invariance_History`.
>
> Update the opinion of each individual by adding the influence and subtracting the invariance.
>
> Store the current opinions in `Opinion_History`.

## Simulation Execution

The extended simulation function is executed to obtain the evolution of opinions, influences, and invariances. These values are then visualized over different time intervals.

## Influence and Invariance Scores

For each time interval:

> Compute the average influence and invariance for each class of individuals.
> Plot these values with respect to the time step.

## Parameters Explanation

**opinions** A list of initial opinions of each individual.

**D** A matrix representing how much each individual influences others.

**class_parameters_list** Parameters associated with each class, including trust towards others, external influence, and invariance.

**class_indices** An array indicating the class each individual belongs to.

**time_steps** The total number of time steps for the simulation.

**transitions**: This is a list that stores all possible types of opinion transitions.

**transition_counts**: This is a dictionary storing the counts of opinion transitions for each class. At initialization, the count for each transition is set to 0.

Transition counts are computed as follows:

(a) Calculate the average opinion for each class between time steps $t-1$ and $t$.
(b) Identify the type of opinion transition between the two time points using the `detect_transition` function.
(c) Update the count for the identified transition.

Every 100 time steps, the counts of opinion transitions for each class are visualized using a bar chart. The transition counts are then reset for the next interval.

# 6. Discussion

**(1) Identifying influential classes**

Figure 2 and Figure 3, The first figure (final opinion distribution) shows that most opinions converge to very small (negative) values; the second figure (change in average opinion over time) shows the change in opinion for each class over time, with some classes showing greater fluctuations in opinion values than others. The third figure (Class Parameters) shows the values of parameters *A*, *B*, and *C* for each class, providing clues as to how these values affect opinion formation.

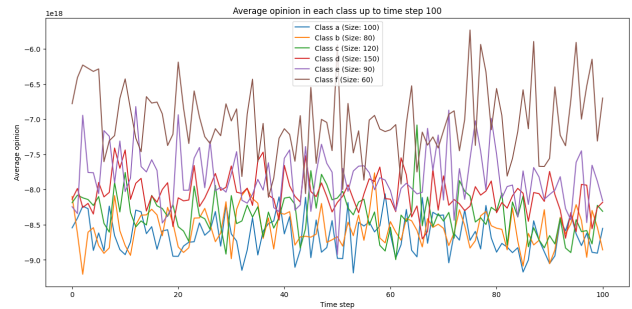

Fig. 2: Simple Opinion Dynamics, $t = 100$

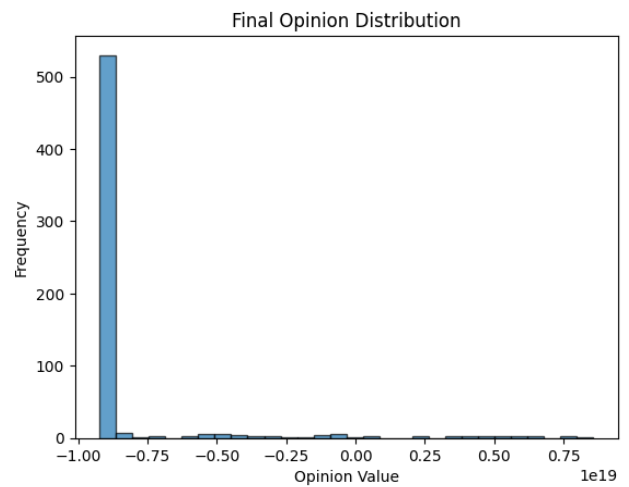

Fig. 3: Final Opinion Distribution

It is important to consider that class a has the highest value of A, which represents the highest confidence in their opinion. Nevertheless, the final opinion distribution converges to very small (negative) values, which suggests that class a is not necessarily the most influential. The fact that the opinion of class d is higher than the other classes over time suggests that class d may have some influence. However, given that the final opinion distribution converges to negative values, it is possible that classes with small values of a, such as class f, have an influence on opinion formation.

### (2) Consideration of social clusters and opinion formation patterns

From the second figure, we can see how the opinions of each class change over time, but none of the classes show clear consistency. The large fluctuations in opinion suggest that there is strong interaction within the social clusters. For example, *a* class with highly fluctuating opinions (e.g. class d) may be more susceptible to external influences (parameter *B*), or it may imply greater dependence on the opinions of other classes (the efficacy C of the invariant field).

If the value of parameter *C* for each class is relatively high (conservative thought), it may mean that opinions tend to be less volatile. In reality, however, opinions in all classes fluctuate significantly, which may indicate that conservatism (*C*) is not sufficient to prevent opinion fluctuation or that external influences (*B*) are strong, indicating that the balance between internal trust (*A*) and external influences (*B*) is important in opinion formation.

### (3) Contributing factors to final opinion convergence

The convergence of final opinions to one implies strong interactions among classes or dominant influence by some classes. It is not clear how the class parameter C contributes to opinion convergence, but the fact that opinions converge to negative values suggests that classes with low *A* values and high *B* values may have a significant influence on opinion formation. This also suggests that the clusters as a whole are more receptive to outside influences.

Social opinion formation is considered to be a complex process in which multiple factors are intricately intertwined in its formation.

### Minorities of Influence (Class *f*)

In real society, what is considered is a minority group with a specific influence. For example, this is the case of a small group of experts, scientists, or activists who have a strong influence on a particular issue. They may be small in number, but have great opinion-forming power due to the persuasiveness and expertise of their opinions (high parameter *B*).

### Groups with high self-trust but isolated (class *a*)

Can represent a group with high self-trust but little influence over other groups or society as a whole. This is the case, for example, in closed communities with strong ideologies or people who hold strongly to certain beliefs and values and hold their own opinions without accepting much outside information.

### Class that is susceptible to outside influences (Class *d*)

This class, whose opinions change dramatically over time, may refer to the general public that is susceptible to media and social media influence. These groups are sensitive to trending topics and news and may frequently change their opinions based on outside information. Society as a whole sees a convergence of opinions.

Societies that ultimately converge in one opinion indicate that they have formed some sort of consensus or that their opinions are shaped by certain strong influences (government, media, social movements). This can be seen, for example, in economic crises, social crises, or when majority and minority opinions converge in policy decisions at the national level.

### (1) Of the six classes *a* − *f*, which class opinion has more influence? The size of class a is the smallest and *f* is the largest.

The value of *A* (confidence in one's opinion) is maximum for a and minimum for f. The value of *B* (external influence) is randomly assigned. The value of *C* (the potency of the invariant field, i.e., conservative thought) is higher in the class where a is high and b is low. Under these conditions, classes a and b are considered the most influential. Class a is the smallest in size but has the most confidence in its own opinion (high value of *A*). This means that members of class a are more likely to maintain their opinions and less likely to be influenced by outside influences. Similarly, class b has confidence in its own opinions and has a reasonably large size, which means that it may play an important role in the formation of society's opinions.

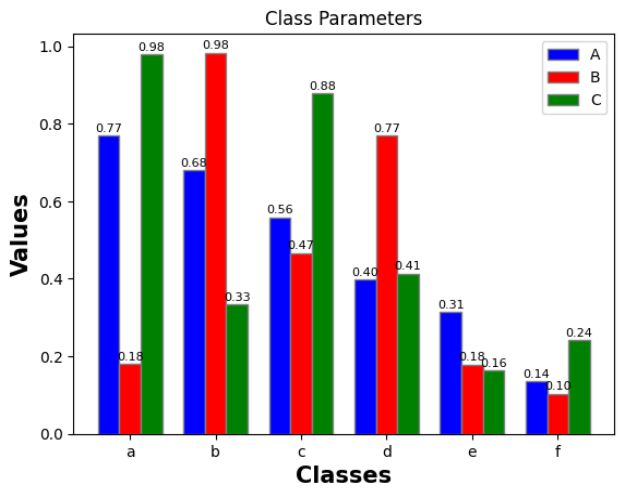

Fig. 4: Class Parameters

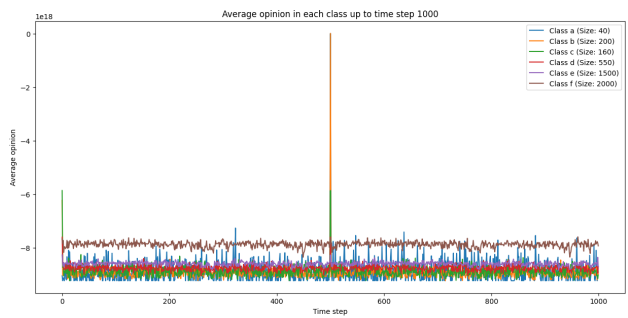

Fig. 5: Average opinion in each class up to time step,Finalsteps

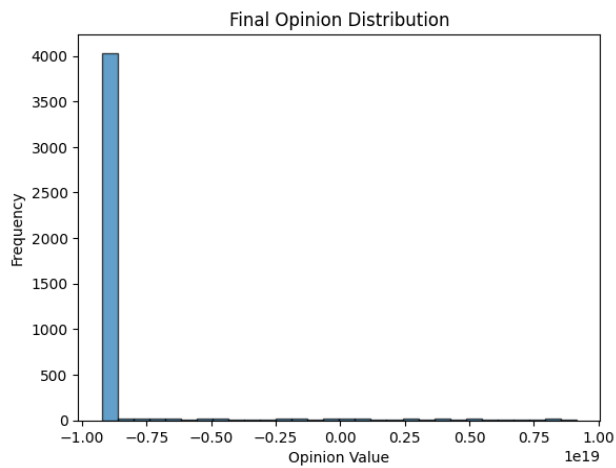

Fig. 6: Final Opinion Distribution)

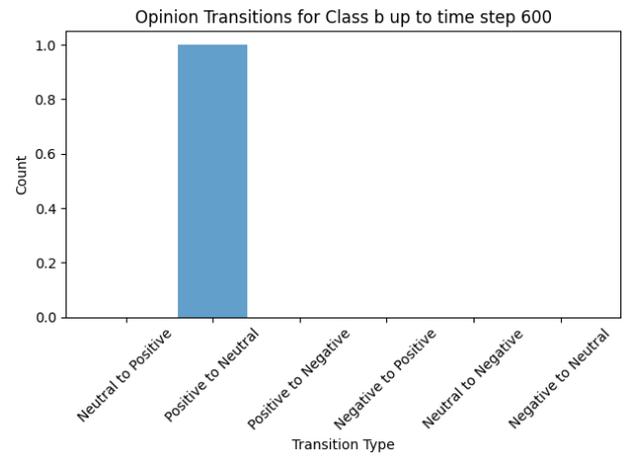

Fig. 7: Opinion in each class up to time step, Transition Type

**(2) Consideration of social clusters and patterns of opinion formation**

The distribution of opinions is ultimately biased to the left, despite the size of class f. This suggests that smaller but more confident groups in their opinions have a greater influence in opinion formation than larger groups. The influence of the invariant field (the value of $C$) indicates that those with particularly conservative opinions are resistant to external change, which may indicate a tendency to retain existing values and traditions rather than innovate in opinion formation. The graph of average opinion shows that the dispersion of opinion is stable over time, indicating that a strong consensus of opinion has formed within the social cluster.

**(3) The final opinions seem to converge to one, but which of the $a - f$ or $A - C$ conditions contributed?**

The distribution of final opinions is biased to the extreme left, which suggests that the small classes a and b with high confidence in their own opinion (value of $A$) played a dominant role in opinion formation. Classes with high invariant field efficacy (value of $C$) may also have contributed to the convergence of opinions. This indicates that conservative values are dominant in the society, implying that resistance to change plays an important role in shaping the eventual convergence of opinions. On the other hand, outside influences ($B$ values) appear to be less important in this process. This indicates that external influences do not have a significant impact on the final opinion formation since the distribution of opinions is very stable over time.

**Influence of Political Extremist Groups**

As a class $a$ (small but confident group), one can think of a small activist group that is politically extreme. This group is a minority in society, but they have very strong beliefs in their views and are very resistant to outside influence. Their beliefs can have a major impact in society and can have a significant influence on political direction. A class $b$ could be a medium-sized group with relatively confident opinions. This might correspond, for example, to a group of lobbyists or activists who are dedicated to a particular policy or philosophy. They play an important role in shaping the direction of public debate.

**Resistance of social and cultural conservatives**

Classes with high $C$ values correspond to conservatives who value traditional values and the existing social order. Their resistance to change can affect the convergence of social opinions and slow the rate at which new ideas and reforms are accepted by society. cRole of Media and External Influences The value of $B$ (external influence) can be thought of as the impact of media, advertising, and world events on the formation of social opinion. However, since external influences do not seem to play as large a role in the final opinion formation in this model, the media influence can be interpreted as being more involved in the reinforcement and diffusion of existing opinions than in the formation of opinions.

**Social Change and Innovation Delay**

The convergence of opinions in this model suggests a tendency for social change and the introduction of innovative ideas to be delayed. Because small, influential and conservative groups dominate the opinion formation process, an argument can be inferred that new policies and social reforms are taking

**longer to be implemented.**

Resistance to technology and innovation The model also shows resistance when it comes to technological advances and social innovations. For example, if a particular group opposes the introduction of a new technology

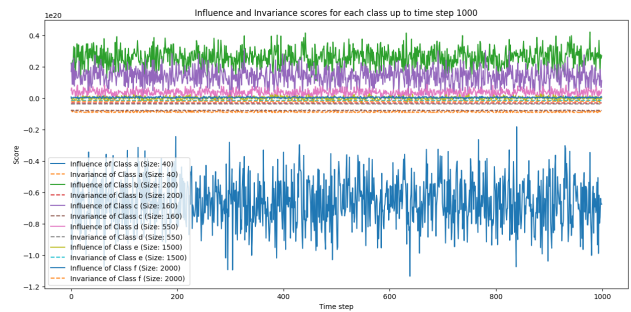

Fig. 8: Influence and Invariance scores for each class up to time step

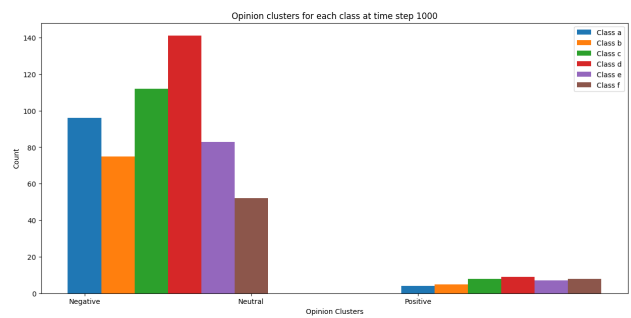

Fig. 9: Opinion clusters for each class at time step

based on a strong identity or tradition, that group may be an important barrier to progress.

# 7. Conclusion

**(1) Which of the six classes $a - f$ has more influential opinions?**

Looking at the second graph, the class with the highest Influence score is Class $e$ (Size 1500), which is almost always positive. This means that this class has the most influence on the other classes; Class $f$ (size 2000) is also influential, but its score is not as high as Class $e$.

**(2) A discussion of the social clusters and patterns of opinion formation that can be assumed from the results of the above graphs and others.**

According to the first graph, many classes are concentrated in Neutral opinions, with much less Negative and Positive opinions. This may indicate that social opinions tend to cluster in the middle, meaning that extreme opinions tend to be in the minority. We can speculate that social clusters tend to reinforce neutral opinions through interaction and external information flows.

**(3) The final opinions seem to converge to one, but please mention which of the a-f or $A - C$ conditions contributed.**

We believe that the most influential class contributed significantly to the convergence of the final opinions. This is Class e and f, with Class *e* in particular appearing to contribute significantly. The $A - C$ conditions indicate that the confidence in Class *a* to *f* ($A$) is high, the external efficacy ($B$) is random, and the influence of the conservative invariant field ($C$) is high when $A$ confidence is high and $B$ is low. In these conditions, Classes $E$ and $F$ play an important role in the final opinion formation because of their high influence and probably also the value of $A$ (confidence in their opinion). The values of $B$ and $C$ are more decisive in opinion formation than the direct influence of the value of $A$. In conclusion, in this simulation, classes with large sizes and high confidence in their own opinions seem to play an important role in opinion formation. In addition, the efficacy of the conservative invariant field ($C$) depends on the values of $A$ and $B$, suggesting that these two parameters are important factors in social opinion formation.

### 7.1 Summary, from Pre-Study

The computational results of this Pre-study hypothetical model infer that each class represents a different opinion group within the society. A class with a high $A$ value (trust in one's own opinion) can be interpreted as having a strong self-trust and a tendency to stick to that opinion. The $B$ values (external influences) are randomly assigned, indicating that each class is more likely to be influenced by different external sources. The value of $C$ (invariant field efficacy, conservatism) is assigned higher to classes with high self-trust and low external influence, which may suggest a group that is more resistant to change and more tradition-oriented.

As an application to real social clusters, we can consider the following:

Classes with high self-reliance and low external influence (e.g. classes *a* and *e*).

These may represent a group of elites, academics, and professionals who hold traditional values and are very confident in their own opinions. They tend to be less influenced by external media and sources and form opinions based on their own expertise and beliefs. Random class of external influences (all classes may fall into this category):.

They may represent the general public and young adults who are more susceptible to social media and news media influences. These groups have characteristics that make their opinions easily influenced by the diversity of information sources. Highly conservative classes (classes with high $C$ values, e.g. classes *a* and *e*).

These may refer to groups that value preserving the existing social order and traditions, such as religious groups, cultural minorities, or groups rooted in a particular region or community. They seek to protect existing beliefs and values rather than embrace new information and change. Given this information, the following cases of discourse are possible

Elite-driven debates: In debates led by academics and experts, opinions based on firm beliefs have a strong influence, which can have a significant impact on the formation of society's opinions. Rise of Populism: Randomly externally influenced classes reflect a modern society with a wide variety of information sources, where populism and mass-congratulatory speech can easily emerge. Traditionalist resistance: resistance to social change may emerge from classes with high conservative invariant field efficacy, and these groups may serve as brakes on social change.

# Aknowlegement

This research is supported by Grant-in-Aid for Scientific Research Project FY 2019-2021, Research Project/Area No. 19K04881, "Construction of a new theory of opinion dynamics that can describe the real picture of society by introducing trust and distrust". It is with great regret that we regret to inform you that the leader of this research project, Prof. Akira Ishii, passed away suddenly in the last term of the project. Prof. Ishii was about to retire from Tottori University, where he was affiliated with at the time. However, he had just presented a new basis in international social physics, complex systems science, and opinion dynamics, and his activities after his retirement were highly anticipated. It is with great regret that we inform you that we have to leave the laboratory. We would like to express our sincere gratitude to all the professors who gave me tremendous support and advice when We encountered major difficulties in the management of the laboratory at that time.

First, Prof. Isamu Okada of Soka University provided valuable comments and suggestions on the formulation of the three-party opinion model in the model of Dr. Nozomi Okano's (FY2022) doctoral dissertation. Prof.Okada also gave us specific suggestions and instructions on the mean-field approximation formula for the three-party opinion model, Prof.Okada's views on the model formula for the social connection rate in consensus building, and his analytical method. We would also like to express our sincere gratitude for your valuable comments on the simulation of time convergence and divergence in the initial conditions of the above model equation, as well as for your many words of encouragement and emotional support to our laboratory.

We would also like to thank Prof.Masaru Furukawa of Tottori University, who coordinated the late Prof.Akira Ishii's laboratory until FY2022, and gave us many valuable comments as an expert in magnetized plasma and positron research.

In particular, we would like to thank Prof.Hidehiro Matsumoto of Media Science Institute, Digital Hollywood University. Prof.Hidehiro Matsumoto is Co-author of this paper, for managing the laboratory and guiding us in the absence of the main researcher, and for his guidance on the elements of the final research that were excessive or insufficient with Prof.Masaru Furukawa.

And in particular, Prof.Masaru Furukawa of Tottori University, who is an expert in theoretical and simulation research on physics and mathematics of continuum with a focus on magnetized plasma, gave us valuable opinions from a new perspective.

His research topics include irregular and perturbed magnetic fields, MHD wave motion and stability in non-uniform plasmas including shear flow, the boundary layer problem in magnetized plasmas, and pseudo-annealing of MHD equilibria with magnetic islands.

We received many comments on our research from new perspectives and suggestions for future research. We believe that Prof.Furukawa's guidance provided us with future challenges and perspectives for this research, which stopped halfway through. We would like to express sincere gratitude to him.

We would like to express my sincere gratitude to M Data Corporation, Prof.Koki Uchiyama of Hotlink Corporation, Prof.Narihiko Yoshida, President of Hit Contents Research Institute, Professor of Digital Hollywood University Graduate School, Hidehiko Oguchi of Perspective Media, Inc. for his valuable views from a political science perspective. And Kosuke Kurokawa of M Data Corporation for his support and comments on our research environment over a long period of time. We would like to express our gratitude to Hidehiko Oguchi of Perspective Media, Inc. for his valuable views from the perspective of political science, as well as for his hints and suggestions on how to build opinion dynamics.

We are also grateful to Prof.Masaru Nishikawa of Tsuda University for his expert opinion on the definition of conditions in international electoral simulations.

We would also like to thank all the Professors of the Faculty of Engineering, Tottori University. And Prof.Takayuki Mizuno of the National Institute of Informatics, Prof.Fujio Toriumi of the University of Tokyo, Prof.Kazutoshi Sasahara of the Tokyo Institute of Technology, Prof.Makoto Mizuno of Meiji University, Prof.Kaoru Endo of Gakushuin University, and Prof.Yuki Yasuda of Kansai University for taking over and supporting the Society for Computational Social Sciences, which the late Prof.Akira Ishii organized, and for their many concerns for the laboratory's operation. We would also like to thank Prof.Takuju Zen of Kochi University of Technology and Prof.Serge Galam of the Institut d'Etudes Politiques de Paris for inviting me to write this paper and the professors provided many suggestions regarding this long-term our research projects.

We also hope to contribute to their further activities and the development of this field. In addition, we would like to express our sincere gratitude to Prof.Sasaki Research Teams for his heartfelt understanding, support, and advice on the content of our research, and for continuing our discussions at a time when the very survival of the research project itself is in jeopardy due to the sudden death of the project leader.

We would also like to express our sincere gratitude to the bereaved Family of Prof.Akira Ishii, who passed away unexpectedly, for their support and comments leading up to the writing of this report. We would like to close this paper with my best wishes for the repose of the soul of Prof.Akira Ishii, the contribution of his research results to society, the development of ongoing basic research and the connection of research results, and the understanding of this research project.